# Enhanced nanosecond electro-optic effect in isotropic and nematic phases of dielectrically negative nematics doped by strongly polar additive

Bing-Xiang Li, Volodymyr Borshch, Hao Wang, Quan Li, Sergij V. Shiyanovskii,

and Oleg D. Lavrentovich[*]

*Liquid Crystal Institute and Chemical Physics Interdisciplinary Program,*

*Kent State University, Kent, OH, 44242, USA*

**Abstract**

An electric field can induce or modify optical birefringence in both the isotropic and nematic phases of liquid crystals (LCs). In the isotropic phase, the electric field induces birefringence with an optical axis along the field. The phenomenon is known as the Kerr effect. In the nematic, the analog of the Kerr effect is the change of existing birefringence through nanosecond electric modification of order parameters (NEMOP) that does not require realignment of the optic axis. The utility of both effects for practical applications is challenged by a relatively weak birefringence induced by the field. We address the issue by adding a non-mesogenic additive 2, 3-dicyano-4-pentyloxyphenyl 4'-pentyloxybenzoate (DPP) with a large transverse dipole moment to mesogenic materials in order to enhance their negative dielectric anisotropy. The DPP doping substantially increases the field-induced birefringence in both NEMOP and Kerr effects, up to 0.02. The doping also slows down the switching processes, but this effect can be compensated by rising working temperatures, if necessary. The enhancement of field induced birefringence by the non-mesogenic dopant paves the way for practical applications of nanosecond electro-optic effects.

[*]olavrent@kent.edu

1. Introduction

Anisotropy of optic and dielectric properties of nematic liquid crystals (NLCs) is utilized in displays, optic shutters, modulators, switches, beam steerers, and other devices [1]. The NLCs possess a uniaxial symmetry and birefringence $\Delta n = n_e - n_o$, where $n_e$ and $n_o$ are the extraordinary and ordinary refractive indices, respectively. The dielectric permittivity of NLCs is also anisotropic, $\Delta \varepsilon = \varepsilon_\parallel - \varepsilon_\perp$, where $\varepsilon_\parallel$ and $\varepsilon_\perp$ are measured along and perpendicularly to the optic axis (director $\hat{n}$) [2, 3]. Externally applied electric fields cause the optical effects in the NLCs through several different mechanisms, such as reorientation of the optic axis known as the Frederiks effect [4-6], modification of the scalar order parameters [2, 7-12], quenching of director fluctuations [2, 13-15], etc. At the beginning of the 80-ies, Yuriy Reznikov et al observed a new effect in which modification of the optical properties of NLCs was caused by the photo-induced molecular conformations [16-20]. The fastest (nanoseconds and tens of nanoseconds) of these mechanisms is the order parameters modification that does not involve director realignment and occurs at the microscopic level [10-12]. This pure scalar order parameters phenomenon is called the "nanosecond electrically modified order parameter" effect, or the NEMOP effect [10-12].

Isotropic fluids can also demonstrate electro-optic response. A strong electric field induces birefringence with the optic axis aligned along the field, which is known as the Kerr effect. The Kerr effect has been studied for isotropic phases of non-mesogenic [21-23] and mesogenic materials [24-31].

NEMOP and Kerr effects are schematically illustrated in Fig.1. In the nematic phase, the optic tensor of NLCs can be considered as an ellipsoid with $\tilde{\varepsilon}_x = \tilde{\varepsilon}_z < \tilde{\varepsilon}_y$ at the field-free state, where $\tilde{\varepsilon}_x, \tilde{\varepsilon}_z$, and $\tilde{\varepsilon}_y$ are the components of optic tensor and the director $\hat{n}$ is along the y axis, see Fig. 1a. As described in Refs.[10-12], the electric field $\mathbf{E}$ applied perpendicularly to $\hat{n}$ in the NLC with $\Delta\varepsilon < 0$ does not realign the director but modifies the scalar order parameters and changes the optic tensor components to the new values $\tilde{\varepsilon}'_x < \tilde{\varepsilon}'_z < \tilde{\varepsilon}'_y$, Fig. 1b. These changes constitutes the essence of the NEMOP effect. The field-induced modification of the optic tensor is fast (at the scale of nanoseconds and tens of nanoseconds), since there is no realignment of the director, and results in clearly detectable field-induced birefringence associated with the changes of the scalar part of the order parameter.

The NEMOP effect occurs within the nematic state of the material. In contrast, Kerr effect is associated with the isotropic phase. In the isotropic phase, in absence of the field, the optic tensor is a sphere, $\tilde{\varepsilon}_x = \tilde{\varepsilon}_y = \tilde{\varepsilon}_z$, Fig. 1c. In materials studied in this work, featuring a strong electric dipole moment perpendicular to the long axis of the molecule, an applied electric field induces negative optical anisotropy, namely, $\tilde{\varepsilon}'_x = \tilde{\varepsilon}'_y > \tilde{\varepsilon}'_z$, Fig. 1d.

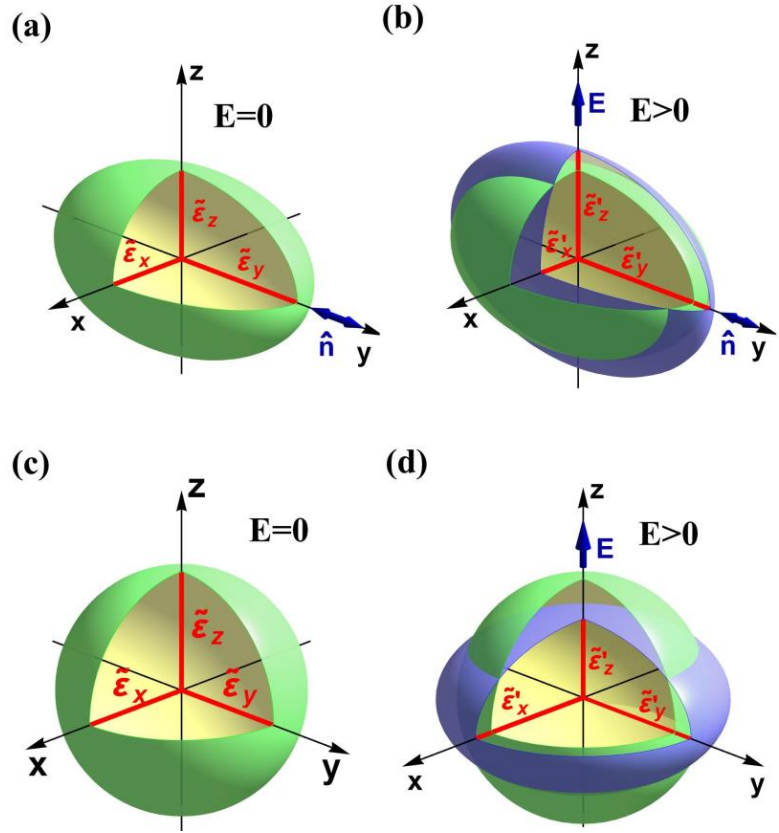

Fig. 1 Illustration of NEMOP effect in the nematic phase (a, b) and Kerr effect in the isotropic phase (c, d) for an NLC with negative static dielectric anisotropy $\Delta\varepsilon < 0$. (a) Ellipsoid of optic tensor in absence of the electric field; the director $\hat{n}$ is along the y axis and $\tilde{\varepsilon}_x = \tilde{\varepsilon}_z < \tilde{\varepsilon}_y$. (b) Modification of the optic tensor by an electric field applied perpendicularly to the director. New components are $\tilde{\varepsilon}'_x < \tilde{\varepsilon}'_z < \tilde{\varepsilon}'_y$. (c) The optic tensor of the NLCs in the isotropic phase is a sphere $\tilde{\varepsilon}_x = \tilde{\varepsilon}_y = \tilde{\varepsilon}_z$. (d) Electric field induces a negative optical anisotropy, $\tilde{\varepsilon}'_x = \tilde{\varepsilon}'_y > \tilde{\varepsilon}'_z$.

The NEMOP and Kerr effects in mesogenic compounds feature different temperature dependencies and amplitudes of the field induced birefringence. As a rule, the Kerr effect shows a strong temperature dependence near the phase transition point [25, 26, 28], while NEMOP features a relatively temperature-independent behavior in the entire range of the nematic phase [32] which might be an advantage is practical applications. Both effects would benefit from the enhancement of the field-induced birefringence which is relatively weak, on the order of 0.01. In this work, we demonstrate that by adding a strongly polar non-mesogenic compound to NLCs with negative dielectric anisotropy, one can increase the field-induced birefringence by several times for both the NEMOP and Kerr effects.

## 2. Experiment

As an additive that enhances the NEMOP and Kerr effect in mesogenic compounds, we used a highly polar compound 2, 3-dicyano-4-pentyloxyphenyl 4'-pentyloxybenzoate (DPP), CAS 67042-21-1 (purchased from UAB Tikslioji Sinteze, Lithuania). The material was purified by chromatography on silica gel (Hexane:Dichloromethane 1:1 volume ratio). An important feature of DPP is a strong electric dipole, about $\mu = 7.6 \text{ Debye}$, oriented perpendicularly to the long axis of the molecule, Fig.2. The dipole was estimated by using Chem 3D Ultra 8.0a in a single molecule approximation, with Molecular Mechanics 2 (MM2) energy minimization procedure and Molecular Orbital Package (MOPAC) properties computation. As shown by Adomenas et al [33], if the DPP molecules could be arranged into a uniaxial nematic phase, such a strong dipole would produce a very strong static dielectric anisotropy, $\Delta\varepsilon \approx -25$. However, despite the fact that the DPP molecules are strongly anisometric, Fig. 2, the material does not exhibit any mesophase [33-35]. It melts into the isotropic phase at $T_{\text{CI}} = 91°C$ [33]. In our work, we thus use DPP as a dopant to the nematic materials with $\Delta\varepsilon < 0$, with the goal to enhance their dielectric anisotropy and birefringence and, as a result, to enhance the amplitudes of the NEMOP and Kerr effects in these mixtures.

In what follows, we characterize the effect of DPP on the NEMOP and Kerr effects for three different nematic compositions with $\Delta\varepsilon < 0$.

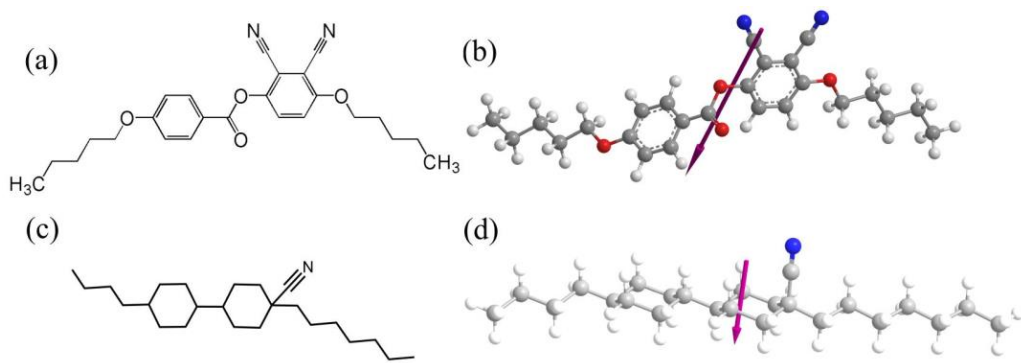

Fig. 2. Chemical structure of (a,b) DPP and (c,d) CCN-47. The arrows in (b) and (d) represent the permanent dipole.

We used NLCs CCN-47, MLC-2080, and HNG715600-100 purchased from their respective vendors, Nematel GmbH (Germany), Merck (USA), and Jiangsu Hecheng Display Technology (China). DPP was added in the same weight proportion of 19% to all three materials. The mixtures were prepared by dissolving the components in 2 ml of chloroform followed by vortexing the solutions. The chloroform solution was slowly evaporated over 12 hours at room temperature. Then the residue was dried in the vacuum at 50°C for 12 hours. The thermal, dielectric and optic properties of studied materials are summarized in Table 1. In all studied materials the DPP doping decreases the clearing temperature $T_{NI}$ and increases the absolute value of negative dielectric anisotropy $\Delta\varepsilon$, determined by the Solatron impedance analyzer using homeotropic and planar nematic cells. Since the chemical structure of CCN-47 is known, we can analyze how strong the effect of adding DPP is. Note that the dipole moment of DPP is about 2 times stronger than the molecular dipole $\mu$ =3.7 Debye [11] of the CCN-47 molecule. By adding 19 wt. % of DPP to CCN-47, the static dielectric anisotropy of the mixture is increased by a factor of 1.6 and the birefringence is enhanced by a factor of 1.3, see Table 1. PolScope [36] measurements of birefringence reveal that the DPP doping increases the intrinsic (field-free) birefringence of the nematic phase of CCN-47 mixture and decreases it for MLC-2080 and HNG715600-100 mixtures, Table 1. The probable reason is that the birefringence of pure CCN-47 is relatively weak, only about 0.03, while un-doped MLC-2080 and HNG7156-100 show a relatively strong birefringence of 0.11-0.15. The effective birefringence of DPP is apparently intermediate between 0.03 and 0.11, but it cannot be measured since DPP does not exhibit a nematic phase.

Table 1. Clearing temperature and dielectric and optic properties ($23°C$) of NLCs materials and their mixtures with DPP.

| Materials | $T_{NI}$ (°C) | $\Delta\varepsilon$ @ 1 kHz | $\Delta n$ @ 589 nm |
|---|---|---|---|
| CCN-47 | 58.5 | −5.1 (40°C) | 0.029 (40°C) |
| DPP:CCN-47 | 53.2 | -8.2 | 0.037 |
| MLC-2080 | 103.1 | −6.4 | 0.113 |
| DPP:MLC-2080 | 90.2 | -9.3 | 0.106 |
| HNG7156-100 | 88.0 | −12.2 | 0.153 |
| DPP:HNG7156-100 | 81.9 | -14.1 | 0.141 |

In order to minimize the *RC*-time, we used glass substrates coated with low-resistivity ($10\,\Omega/\text{sq}$) indium tin oxide (ITO) electrodes of a small area, $2\times 2\,\text{mm}^2$. The cells were assembled from two parallel substrates separated by spherical spacers of the diameter $4.4-6.4\,\mu\text{m}$. The inner surface of the test cells were spin-coated with polyimide PI-2555 (HD MicroSystems) and rubbed unidirectionally. The temperature of the cells was controlled with an accuracy of $0.1\,°C$ by LTS350 hot stage (Linkam Scientific Instruments) and Linkam TMS94 controller.

The optical scheme to characterize the optical response in the NEMOP and Kerr effect is shown in Fig.3a. It is described in details in Refs. [10-12, 31]. Here we mentioned that the scheme implies an oblique incidence and particular direction of linear polarization ($45°$ with respect to the incidence plane) of the probing laser beam (He-Ne laser, $\lambda = 632.8\,\text{nm}$) in order to eliminate the contribution of director fluctuations to the optical response in the NEMOP effect. The refractive index of the prisms used for oblique incidence, $n_g = 1.52$, was close to the refractive indices of the glass substrates and NLCs. The beam traveled the optical stack in the following order: the polarizer, NLC cell, Soleil-Babinet compensator, and analyzer [10-12]. The transmitted light intensity was measured by a photodetector TIA-525 (Terahertz Technologies, response time < 1 ns). Voltage pulses of amplitude up to 1 kV with nanosecond rise and fall fronts were generated by a pulse generator HV 1000 (Direct Energy, Inc.), Fig. 3b. In all the experiments, the duration of the applied square voltage pulses is ~ 400 ns. In Fig.3b, this pulse extends from $t \approx 100$ ns to $t \approx 500$ ns. Note that the switching on and switching off times of the pulse are within 5 ns, which is much shorter than the characteristic response times of the nematics. The profiles of voltage $U(t)$ and optical response were determined with an oscilloscope Tektronix TDS 2014 (sampling rate 1 GSample/s).

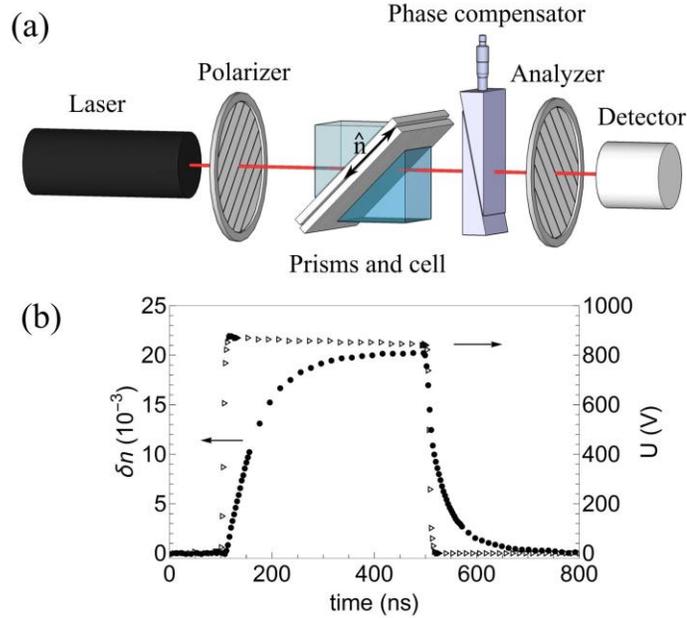

Fig. 3. (a) Experimental setup; (b) Dynamics of field-induced birefringence (filled circles) for DPP:HNG715600-100 mixture in response to the applied voltage (triangles) in a cell of thickness $d = 4.4\,\mu\text{m}$ at $23°\text{C}$.

To describe the dynamics of the system response, we introduce the rise time $\tau_r$ defined as the time taken by the optical response signal to change from 10% to 90% of its maximum intensity and the fall time $\tau_f$, defined as the time during which the optical response signal decreases from 90% to 10%. We estimate the errors below 5% for the field-induced birefringence and 10% for the rise and fall times.

### 3. Results and discussions

The electro-optic response of all three DPP doped NLCs shows a significant enhancement of the field-induced birefringence for both the NEMOP and Kerr effects. The DPP doping of CCN-47 results in the largest amplitude increase for both effects, Fig.4a,b.

Let us analyze the effect of DPP on the field-induced NEMOP birefringence in its mixture with CCN-47. When the comparison is performed for the same absolute temperature (43-44 °C), the data show a factor of 4 of birefringence enhancement in the mixture, Fig. 4a,b. One can also compare the performance of the pure CCN-47 and its mixture with DPP at the same *reduced* temperature $T^*$ defined as $T/T_{NI}$, see e.g. [37], where $T_{NI}$ is the nematic-to-isotropic phase transition, equal 58.5 °C for CCN-47 and 53.2 °C for its mixture with DPP, Table 1. For the same $T^* = 0.972$ ($T = 49.1$ °C for CCN-47 and $T = 44.0$ °C for the mixture), the field induced birefringence of pure CCN-47 is $0.5 \times 10^{-3}$, while for the mixture it is $1.7 \times 10^{-3}$, for the same field amplitude $E = 1.1 \times 10^8$ V/m.

The DPP doping of CCN-47 also enhances the amplitude of the Kerr effect, by about 50%, Fig.4a,b. At the same time the doping slows down the rise and fall processes for both NEMOP and Kerr effects, Fig.4c,d. Note that the rise times decrease at higher applied fields, Fig. 4c.

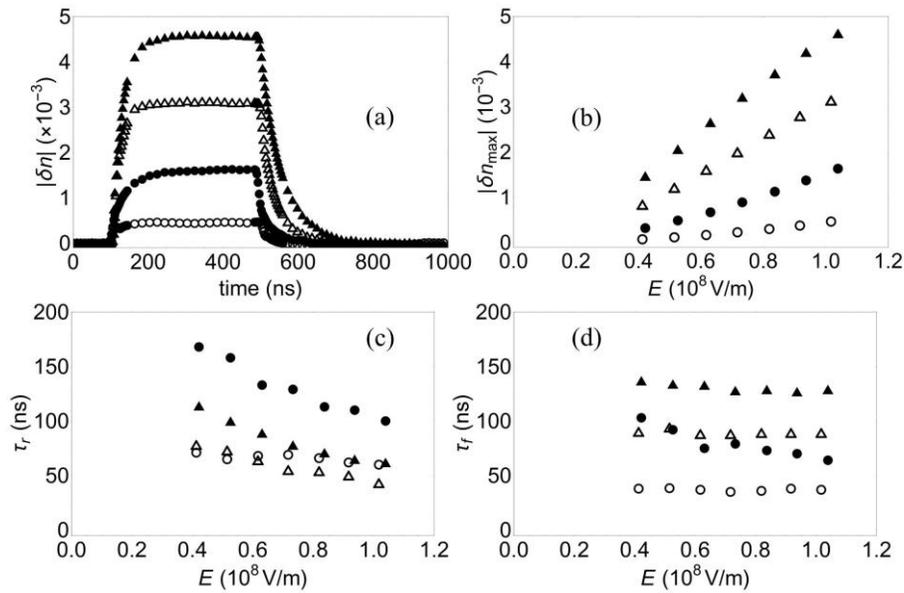

Fig.4. The NEMOP effect, circles (43°C ○, 44°C ●) and the Kerr effect, triangles (68.5°C △, 63°C ▲) in pure CCN-47 (open symbols ○, △) and in DPP:CCN-47 mixture (filled symbols ●, ▲): (a) dynamics of the field-induced birefringence $\delta n$ at $E = 1.1 \times 10^8$ V/m; electric field dependencies of (b) peak value of field-induced birefringence $\delta n_{max}$: (c) rise time $\tau_r$; and (d) fall time $\tau_f$.

The DPP doping of MLC-2080 yields a substantial enhancement of birefringence, about 2.5 times for the NEMOP effect, Fig.5a,b. Although the dopant also leads to a significant increase of the rise time, Fig.5c, the DPP:MLC-2080 is still the fastest mixture among the three, with a typical rise time of 100 ns and the fall time about 40 ns, Fig.5c,d.

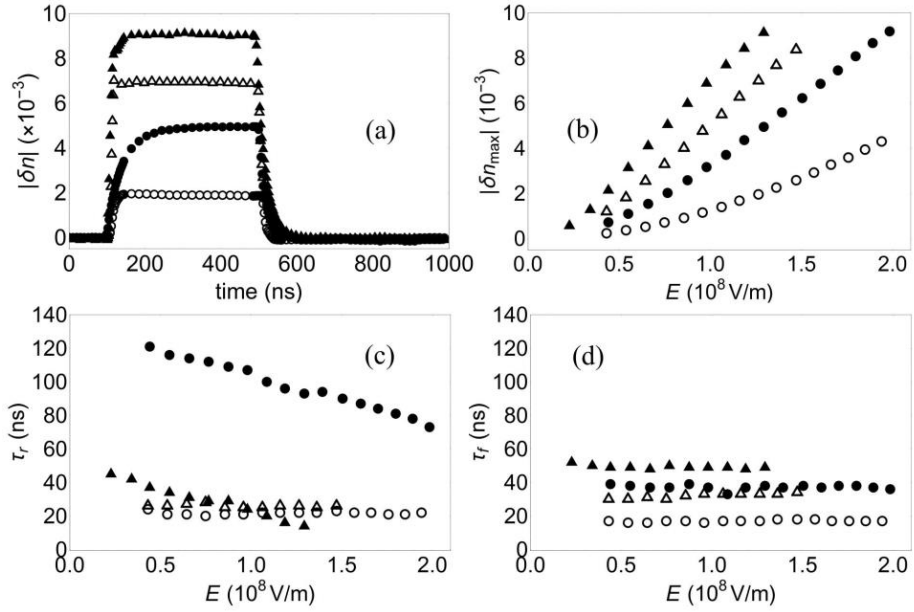

Fig.5. The NEMOP effect, circles ( 23°C $\bigcirc$, 23°C $\bullet$ ) and the Kerr effect, triangles ( 113°C $\triangle$, 100°C $\blacktriangle$ ) in pure MLC-2080 (open symbols) and in DPP:MLC-2080 mixture (filled symbols): (a) dynamics of $\delta n$ at $E = 1.3 \times 10^8$ V/m; the electric field dependencies of (b) $\delta n_{max}$, (c) rise time $\tau_r$; (d) fall time $\tau_f$.

The mixture DPP:HNG715600-100 exhibits the largest birefringence enhancement for both effects, Fig.6a,b. We also observed that the DPP doping increases the rise and fall times for both effects, but the slowing down can be compensated by rising the working temperature, compare the data for the NEMOP effect presented in Figs.6c,d for the temperatures 23°C and 50°C.

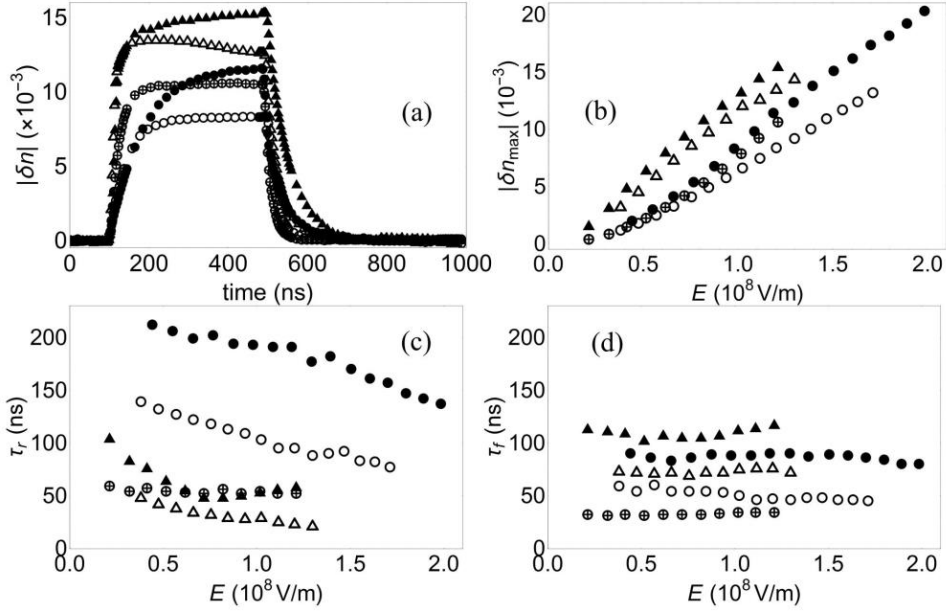

Fig.6. The NEMOP effect, circles ( 23°C ○, 23°C ●, 50°C ⊕) and the Kerr effect, triangles ( 93°C △, 87°C ▲) in pure HNG715600-100 (open symbols) and in DPP:HNG715600-100 mixture (filled and crossed symbols): (a) dynamics of the $\delta n$ at $E = 1.2 \times 10^8$ V/m, and the electric field dependencies of (b) $\delta n_{max}$, (c) rise time $\tau_r$ and (d) fall time $\tau_f$.

### 4. Conclusions

To summarize, the NEMOP effect and the Kerr effect exhibit nanosecond electro-optic response that is appealing for applications where the speed of switching is of prime importance. However, to make these effects really practical, one needs to enhance the amplitude of the field-induced birefringence. In this work, we demonstrate that doping NLCs with strongly polar additives is an effective approach to increase the field-induced birefringence of both effects.

In the NEMOP effect, the maximum field-induced birefringence of the studied doped materials is $\delta n_{max} = 0.02$. The typical response times are within hundreds of nanoseconds and often faster. The achieved value of field-induced birefringence is sufficient to switch between two orthogonal linear polarization states and thus produce light intensity modulation from zero to maximum when the NLC cell is placed between two crossed polarizers. To see that such a switching is possible, one can estimate the total optic retardance of a cell with the thickness $d = 6\,\mu\text{m}$ in which one electrode represents a mirror and for which the light incidence angle is 45º. With the data above and with $\delta n = 0.02$, the total optical retardance is $\Gamma = 2\delta n\, d\sqrt{2} = 340$ nm, which is larger than the half-wavelength of a He-Ne laser. Designing the NLC cells in such a way that the light beam experiences several reflections, one can achieve phase retardances up to $1\,\mu\text{m}$ [38]. In the nematic case, the NEMOP effect is practically temperature independent making it promising for applications.

For the Kerr effect in the isotropic phase, the maximum field-induced birefringence is $\delta n_{max} = 0.015$ which is 1.3 times higher than the optical response in the nematic phase at the same electric field. In addition, the optical response shows a strong enhancement at temperatures near the isotropic-to-nematic phase transition.

Because of their different physical nature and different dependencies on temperature, the NEMOP and Kerr effects can complement each other in terms of applications. In particular, the NEMOP effect could be used in ultrafast NLC applications that can tolerate reasonable temperature variations. On the other hand, the Kerr effect can provide larger birefringence in devices that can provide a strict temperature control. The studies presented in this work suggest that the synthesis of other strongly polar compounds as additives to NLC materials would be a promising avenue for the enhancement of nanosecond electro-optical NEMOP and Kerr effects.

**Acknowledgements**

The work was supported by National Science Foundation (NSF) grant DMR-1410378 and IIP-1500204.